\documentclass{vldb}
\usepackage{graphicx}
\usepackage{balance}
\usepackage{epsfig}
\usepackage{endnotes}
\usepackage{tabularx}
\usepackage{multirow}
\usepackage{listings}
\usepackage{xspace}
\usepackage{xcolor}
\usepackage{textcomp}
\usepackage{placeins}
\usepackage{array}
\usepackage{multirow}
\usepackage{courier}
\usepackage{color}
\usepackage{paralist}
\usepackage{url}
\usepackage{subcaption}
\usepackage{floatrow}
\usepackage{appendix}
\usepackage{multirow}

\newcommand{\ttcode}{\ttfamily\small}

\newlength{\LL} 
\newcolumntype{L}{>{\raggedright\arraybackslash}X} 
\newcolumntype{C}{>{\centering\arraybackslash}X}   

\begin{document}

\date{}

\title{
A Fast Lightweight Time-Series Store for IoT Data
}



\numberofauthors{2}
\author {
\alignauthor
Daniel G. Waddington\\
       \affaddr{Samsung Research America}\\
       \affaddr{Mountain View, CA}\\
       \email{daniel.waddington@acm.org}
\alignauthor
Changhui Lin\\
       \affaddr{Samsung Research America}\\
       \affaddr{Mountain View, CA}\\
       \email{changhui.lin@samsung.com}
}

\maketitle

\begin{abstract}

With the advent of the Internet-of-Things (IoT), handling large
volumes of time-series data has become a growing concern.  Data,
generated from millions of Internet-connected sensors, will drive new
IoT applications and services.  A key requirement is the ability to
aggregate, preprocess, index, store and analyze data with minimal
latency so that time-to-insight can be reduced.  In the future, we
expect real-time data collection and analysis to be performed both on
small devices (e.g., in hubs and appliances) as well in server-based
infrastructure.  The ability to localize sensitive data to the home,
and thus preserve privacy, is a key driver for small-device
deployment.

In this paper, we present an efficient architecture for time-series
data management that provides a high data ingestion rate, while still
being sufficiently lightweight that it can be deployed in embedded
environments or small virtual machines.  Our solution strives to
minimize overhead and explores what can be done without complex
indexing schemes that typically, for performance reasons, must be held
in main memory.  We combine a simple in-memory hierarchical index,
log-structured store and in-flight sort, with a high-performance data
pipeline architecture that is optimized for multicore platforms.  We
show that our solution is able to handle streaming insertions at over
4 million records per second (on a single x86 server) while still
retaining SQL query performance better than or comparable to existing
RDBMS.

\end{abstract}


\section{Introduction}

By 2020, analysts predict that there will be over 25 billion connected
devices in the Internet-of-Things (IoT).  Together, these devices will
generate unprecedented volumes of data that must, in order to create
value, be efficiently indexed, stored, queried and analyzed.  A single
aspect that ties all of this data together is that it is {\it
  time-series}.  Time is either evident as an explicit dimension in
the data (e.g., a sensor generated time-stamp) or implicit by the time
at which a data sample reaches an aggregation node (e.g., a hub or
server).

Existing databases and storage systems predominantly handle
time-series data no differently from other data.  There does exist a
notion of time-series database, some that are based on RDBMS and some
that based on NoSQL type architectures.  While APIs and data types may
be provided specifically for time, the underlying storage and indexing
mechanisms are no different.  As a consequence of generality, existing
RDBMS solutions are typically limited to ingress streams of 200-300K
RPS on a single commodity server and 25-30K RPS on an embedded
platform.  Their performance is limited primarily because of the
complexity introduced by indexing with
B+-trees~\cite{Elmasri:1999:FDS:554501} or LSM
trees~\cite{O'Neil96thelog-structured} as well as the maintenance of
locks and state for transactional processing.

However, a key strength of RDBMS solutions is their ability to support
advanced queries through SQL support~\cite{Groff:2009:SCR:1594046}.
SQL provides a powerful language for performing arbitrary queries that
require filtering (selection), manipulation (e.g., data conversion),
projection (e.g., aggregation), and joins.  SQL also provides a
standardized and industry accepted interface to which analytic
solutions (e.g., Apache Spark~\cite{Zaharia:2010:SCC:1863103.1863113},
Tableau~\cite{tableau}, SAS~\cite{sas}, SAP~\cite{sap}) and
applications can be easily integrated.

We believe that this type of complex query is a key enabler for future
IoT applications that derive value by creating insight from data.  By
digitally enabling a product through sensor augmentation, massive
volumes of data can be collected about operation, wearing and
environmental conditions.  Combining this with analytics that can
quickly slice-and-dice the data, enables new hybrid product-service
models to be realized (e.g., preventative maintenance, energy
optimization).

While most database solutions provide SQL access to data, they
inherently cannot support continuous ingest of large-volumes of
streaming data.  Those solutions that do provide higher ingest rate
capabilities typically make extensive use of main memory to hold
indexing information. 

\vspace{2mm} 

The focus of this work is the development of a lightweight and
memory-efficient architecture that supports an order-of-magnitude
higher ingest rate over existing RDBMS and time-series database
solutions.  We do this while still providing both the flexibility and
power of SQL, and a minimal run-time memory overhead so as to be
suitable for deployment in embedded or resource-limited environments.

Our solution, herein termed Lightweight Time-Series Store (LTSS), is
domain-specific in that it is designed to leverage the
characteristics of time-series data in an IoT application context.

Specifically, LTSS is designed with the following assumptions
in mind:

\begin{enumerate}[i]

\item {\it Data is highly-ordered} - Time advances and thus data is
  inherently in order. Network routing and imprecise clock
  synchronization (i.e., inability to have a true global clock) can
  cause out-of-order data, but this is typically limited to a finite
  time window (samples outside of such a window can be considered
  invalid and not useful).
\item {\it Data is immutable} - Each data element represents a sample
  in time that may or may not be subsumed by a subsequent sample, but
  in itself represents an unchangeable history.  Applications
  generally summarize and aggregate data to create useful insight.
\item {\it Both epoch and calendar-based reasoning are necessary for
  analytics} - While epoch-based time are useful, alone they are
  insufficient.  Many analyses require reasoning in terms of
  human-understandable concepts (e.g., hours, days, months). For example,
  queries such as ``Is the current value the highest for the month?'' are
  typical.
\item {\it Data can be lost} - Most data is collected over unreliable
  wireless network links and through congested IP routers.
  Sensor-based data sources do not, in the most part, have the ability
  to perform re-transmission due to limited energy and memory
  capacities.  IoT applications are predominantly resilient to data
  loss.

\end{enumerate}

\subsection{Deployment Requirements}

Our architecture focuses on indexing and query at the {\it ingestion
  node}.  The ingestion node is the first ``point of contact'' for
data being collected from sensors distributed across a network.  Data
is received either directly from the sensor (e.g., where a sensor has
direct Internet connectivity) or from a hub (e.g., that aggregates
across multiple non-IP sensors). Its role is to receive streaming
data, which is then preprocessed, stored and indexed according to one
or more time fields, and made accessible through read-only query interfaces.

The ingestion node is responsible for the management of ``fresh''
data.  It has limited storage resources and can only store data
for a given period of time.  Old data is either off-loaded
to cold-storage (e.g., in the cloud) or erased.

An ingestion node implementation may take multiple forms depending on
the deployment context.  For example, in the home it may be realized
in one or more hub devices or embedded into existing consumer
electronics (e.g., a SmartTV). At the network edge, the ingestion
node may be deployed on commodity x86 servers or cloud infrastructure.
We do not envisage the deployment of ingestion nodes at the cloud
back-end due to the cost of back-hauling data.

For our work we are particularly focused on in-home and edge deployments.  We
define the following requirements accordingly:

\begin{enumerate}[i.]

\item {\it High-Volume Data Rates} - Support for continuous
  aggregation and summarization of large volumes of streaming data resulting from an
  increasing number of deployed sensors with higher data generation
  rates (i.e., increased sampling frequencies).

\item {\it Flexible Memory Model} - Tailorability of main memory
  overhead according to available resources. Less than 32MB minimum
  memory footprint in an embedded environment.


\item {\it Low-Latency for Real-time Applications} - Freshness of data
  (i.e., the time between the data being received and it being
  accessible to queries) should be in the order of 200-400ms (latency
  deemed necessary for interactive applications).

\item {\it SQL-based Query Interface} - The system should support 
  SQL in order to easily integrate with existing platforms and applications
  while also providing query flexibility to support complex 
  operations and joins across multiple data sources.


\item {\it Data Aging and Resource Re-cycling} - Data should be stored
  using a fixed set of resources (disk,memory).  Aging data should
  override new data as resources are exhausted.

 
\end{enumerate}

\subsection{Outline}

The rest of this paper is organized as follows.  Section 2 presents
some background and related work, and positions our own work in the
field of time-series/streaming databases. In Section 3, we present our
architecture and provide details on how data from continuous streams is
ordered, indexed and stored, and how it is integrated with an SQL
front-end.  Section 4 describes details around the prototype
implementation.  Next, Section 5, presents our evaluation based on
three real-world data sets with representative queries and also
discuss our testing methodology. Finally, in Section 6 we offer our
conclusions.

\begin{table*}
\begin{center}
\begin{tabularx}{1.0\linewidth}{|
>{\setlength{\hsize}{.12\hsize}\raggedright\scriptsize}X|
>{\setlength{\hsize}{.06\hsize}\raggedright\scriptsize}X|
>{\setlength{\hsize}{.14\hsize}\raggedright\scriptsize}X|
>{\setlength{\hsize}{.08\hsize}\raggedright\scriptsize}X|
>{\setlength{\hsize}{.07\hsize}\raggedright\scriptsize}X|
>{\setlength{\hsize}{.09\hsize}\raggedright\scriptsize}X|
>{\setlength{\hsize}{.1\hsize}\raggedright\scriptsize}X|
>{\setlength{\hsize}{.14\hsize}\raggedright\scriptsize}X|
>{\setlength{\hsize}{.2\hsize}\raggedright\arraybackslash\scriptsize}X|}\hline

{\bf Name} & {\bf Impl. Lang.} & {\bf Interface} & {\bf SQL Ad-hoc Queries} & {\bf Cont. Queries} & {\bf License} & {\bf Data Store} & {\bf Data Struct.} & {\bf Advertised Ingest Rate (Per-Node)} \\\hline

Akumuli      & C++   & JSON          & N       & N & Open        & Local         & Log                    & 200K RPS \\\hline
Bolt         & C\#   & C\# API       & N       & N & Proprietary & Distributed   & LSM                    & 100K RPS \\\hline 
Druid        & Java  & HTTP/REST     & N       & Y & Open        & Distributed   & Columnar               & 10K RPS    \\\hline
IBM Informix & -     & REST/SQL/JSON & Y       & N & Commercial  & Local         & unknown                & 420K RPS \\\hline
InfiniFlux   & C++   & SQL           & Y       & Y & Commercial  & Local         & Columnar               & 2M RPS \\\hline
InfluxDB     & Java  & HTTP/REST     & Partial & Y & Open        & Local         & LSM                    & 100-150K RPS \\\hline
OpenTSDB     & Java  & HTTP/CLI      & N       & N & Open        & Distributed   & HBase/Cassandra        & 100K RPS \\\hline
ParStream    & C++   & JDBC/SQL      & Y       & N & Commercial  & Distributed   & Columnar               & 500K RPS \\\hline
PipelineDB   & C++   & SQL           & Y       & Y & Open        & Local         & B-tree                 & 500K RPS \\\hline
Prometheus   & Go    & HTTP/JSON     & Y       & N & Open        & Local         & LSM                    & unknown \\\hline
\hline
LTSS          & C++   & SQL           & Y       & Partial & Proprietary & Local         & Log                    & 4M+ RPS \\\hline

\end{tabularx}
\end{center}
\caption{Popular Time-Series Databases.}
\label{tbl:comp}
\end{table*}

\section{Related Work} \label{sec:related-work}

There currently exists a broad range of time-series databases,
both commercially and as open source.  These databases provide
additional capabilities that are specific to streaming data ingest and
time-oriented queries.  With the advent of IoT, these databases
vendors and developers are touting support for store and analysis of large
volumes of sensor data. The specific features of each are also varied.
Table~\ref{tbl:comp} lists some of the more popular time-series
database solutions and highlights their different feature sets.


The work in this paper focuses on the following key capabilities:
\begin{itemize}
\item High ingest rates
\item Support for an SQL interface
\item Local write-append data storage
\end{itemize}

Of the technologies listed in Table~\ref{tbl:comp}, PipelineDB,
ParStream, IBM Informix and InfiniFlux meet this criteria and are
thus most relevant to our work.  However, they are not designed
with embedded and other resource constrained environments in mind.

Pipeline DB~\cite{pipelinedb}, an open source project primarily
developed by Usman Masood, is aimed at performing SQL queries on
streaming data.  Specifically, it is designed to excel at SQL queries
that reduce the cardinality of streaming data sets through aggregation
and summarization.  Raw data is discarded once it has been read by the
continuous queries that need to read it. Pipeline DB provides full SQL
support and provides additional constructs to support queries on
sliding windows and continuous materialization.  As with our own
philosophy, Pipeline DB aims to eliminate an explicit extract-transform-load
stage from the data analytics pipeline.  The system is based on, and
fully compatible with, PostgreSQL 9.4, and therefore uses B-tree/hash
based indexing.  Pipeline DB uses PostgreSQL's {\ttcode INSERT} and
{\ttcode COPY} constructs for writing data. In the long term, the
Pipeline DB team hope to factor out the solution as a PostgreSQL
extension, suggesting that they do not plan to change the storage
internals.

ParStream~\cite{parstream,parstreamwp}, recently acquired by Cisco, is
a broader analytics platform aimed at IoT (originally ParStream was
touted as a real-time database for big data analytics).  Their key
differentiators are fast ingest, edge analytics and the ability to
create real-time insights through concurrent fast queries.  The
ParStream architecture leverages both in-memory processing (for
indices and data caching) and disk-based storage.  The system has been
explicitly designed to exploit multicore architectures by the use of
parallel processing.  The crux of the ParStream DB indexing is their
High Performance Compressed Index (HPCI), which is currently in patent
application~\cite{parstream:patent}.  HPCI is a compressed bitmap
index on which queries can be directly executed.  This both reduces
the time to scan records and the memory footprint for the index.
Unlike LTSS, the ParStream architecture separates out {\it Import} and
{\it Query} Nodes.  This allows them to alter the ratio of query and
ingest resources.  Import Nodes, basically perform an
extract-transform-load stage, convert incoming data into an internal
data partition format.  It is possible to transform data based on SQL
statements executed in the Import Nodes.  ParStream provides SQL over
JDBC/OBDC, as well as a C++ interface.  The system also supports
cluster-based deployment and replication, which LTSS does not.

InfiniFlux~\cite{infiniflux} is another commercial time-series
database solution.  It is a columnar DBMS tailored to processing
time-series ``machine data'' at high ingest speeds (millions of
records per second on a single node). Like ParStream, InfiniFlux uses
an in-memory bitmap based index to manage the data.  Columns are set
up for each potential value and bit vectors used to indicate the value
for a given row.  This provides significant compression for data
sources whose value ranges are limited (e.g., sensor data ranging for
0-255).  InfiniFlux is designed to achieve maximum analytical
performance by storing records in a hybrid arrangement of column-oriented
coupled with time-partitioned.  InfiniFlux advertises an ingest
rate of around 2M records per second.

Finally, IBM Informix, a broader commercial RDBMS suite, provides a
specific solution for time series data known as {\it Informix
  TimeSeries}~\cite{informix}.  At the core of ITS is the DataBlade
extensibility framework, which allows the user to customize and
enhance types and functions in the database system.  In essence, the
TimeSeries enhancements provide additional data types and compressed
storage schemes.  Similar to the LTSS composite time type (see
Section~\ref{sec:ctime}), Informix TimeSeries provides a {\ttcode
  CalendarPattern} data type that allows intervals based on calendar
fields (e.g., second, minute, hour) to be easily defined.  It also
provides other time-specific data types for regular and irregular
time-series (grouping together rows ordered by time stamp). Time-series
data can be stored in rows (when less than 1500 bytes) or in
{\it containers} that store the data outside of the database tables.
These, like LTSS, are contiguous append-only logs.

All of these solutions are primarily designed with server deployment
in mind.  They intentionally leverage a combination of storage and
(large) main memory to accelerate system performance.  LTSS on the
other hand, makes minimal use of main memory, but is designed to exploit
random access performance of new non-volatile RAM storage technologies.

\section{Architecture} \label{sec:arch}

The LTSS system is designed around a component-based data processing
architecture that allows flexible configuration of real-time compute
and storage.  

\begin{figure}[h]
\centering
\includegraphics[width=1.0\columnwidth]{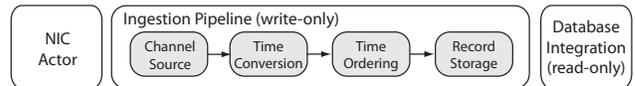}
\caption{LTSS High-Level View}
\label{fig:arch_tihlv}
\end{figure}

The LTSS ingestion pipeline has four main
components (see Figure~\ref{fig:arch_tihlv}):

\begin{itemize}

\item {\it Channel Source} - The Channel Source component has an
  active thread that consumes data from the NIC actor (via a data
  plane channel). It simply takes each record and passes it on through
  a down-stream call.

\item {\it Time Conversion} - Conventional UNIX epoch (or higher
  precision) time stamps are converted to a concise calendar based
  ``composite'' time (see Section~\ref{sec:ctime}).  This allows the
  system to accelerate data selection based on calendar patterns (e.g.,
  every first Tuesday in the month).  Note that this conversion is 
  performed during the data ingest.

\item {\it Time Ordering} - The Time Ordering component ensures that
  ingress packets are ordered according to time.  Out of order records
  may arise from lack of global clock synchronization, Internet
  routing, and OS scheduling. Records are ordered through an insertion
  sort.

\item {\it Record Store} - The final stage of the ingestion pipeline
  is the Record Store.  The Record Store provides an interface for
  write-append, time-ordered storage of records.  Data is stored as a
  log with in-memory metadata to accelerate look up.  We do 
  not currently do any compression on the stored data.

\end{itemize}

The query side (read-only) runs as a separate process that interacts with
the pipeline through shared memory.

\subsection{Composite Time Representation}
\label{sec:ctime}

A calendar-based, {\it composite time} type ({\ttcode ctime\_t}), with
microsecond precision, is used to represent indexed time columns (this
sub-second precision is necessary for IoT applications).  The {\ttcode
  ctime\_t} type uses a compact bit field representation totaling only
8 bytes (see below).  This is considerably more memory efficient than
the POSIX {\ttcode tm} structure that requires 36 bytes.
Calendar-based field extraction is more efficient with composite time
than using a basic time stamp.
 
{\scriptsize
\begin{verbatim}
 unsigned usec:     20;  // range 0-999999
 unsigned sec:      6;   // 0-59
 unsigned min:      6;   // 0-59
 unsigned hour:     5;   // 0-23
 unsigned wday:     3;   // 0-6
 unsigned day:      5;   // 0-31
 unsigned month:    4;   // 0-11
 unsigned year:     5;   // 2000 up to 2031
 unsigned timezone: 5;   // 0-23
 unsigned pm:       1;   // 0-1
 unsigned dls:      1;   // 0-1
 unsigned reserved: 3;
\end{verbatim}
}

All time columns, on which query constraints can be expressed, are
represented with the {\ttcode ctime\_t} type.  Standard converters,
that are executed in the pipeline, are implemented for epoch and ISO
8601 string-based time formats.  Records stored in the LTSS must have
at least one composite time field.  The current implementation only
supports constraints on one instance of composite time. 

\subsection{In-flight Record Ordering}
\begin{figure}
\centering
\includegraphics[width=0.98\columnwidth]{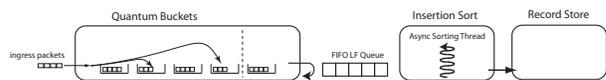}
\caption{Time Ordering Architecture}
\label{fig:arch_toa}
\end{figure}

Ingress records are allocated to ``quantum buckets'' (see
Figure~\ref{fig:arch_toa}).  Each bucket represents a relative time
window.  The target bucket for a given record is simply calculated by
rounding down the record's time stamp (epoch or composite).  Buckets
are created on-demand providing that no future bucket (i.e. with a
greater time stamp) has already been created and closed (a record
attempting to create such a bucket is considered ``delinquent'' and is
dropped).  Buckets are arranged in an ordered queue and expired from
the tail according to time.  The duration of the quantum, and thus
the length of the wait in the queue before sorting and closing,
implicitly defines the freshness of data.  This attribute is
configurable.

When the Time Ordering component expires a bucket, it asynchronously
passes the set of records for the quantum to a sorting thread.  This
performs an insertion sort on the records before passing them to the
Record Store.  We use an insertion sort because of its good
performance in sorting nearly ordered lists with typically a small
number of places an element can be out of
order~\cite{Cook:1980:BSA:359024.359026}.

\subsection{In-Memory Directory Index} 
\begin{figure}
\centering
\includegraphics[width=0.98\columnwidth]{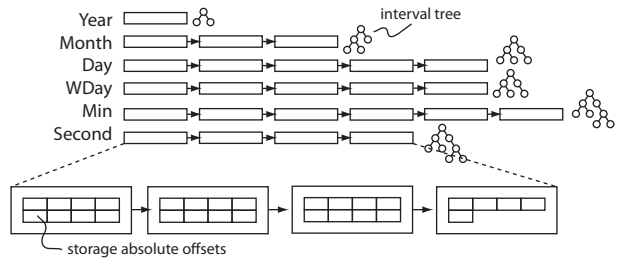}
\caption{Record Store Metadata}
\label{fig:rsmeta}
\end{figure}

The Record Store uses a shared memory directory-style index to
accelerate look up performance.  A dynamic list of absolute byte
offsets is maintained for each composite time field down to second
granularity (see Figure~\ref{fig:rsmeta}).  Each list maintains
offsets to the starting point for {\it every} instance of the given
field.  For example, the Year list contains a single offset for each
year the data covers.  The Month list contains an offset for each
month covered and so on and so forth.  Interval
trees~\cite{Cormen:2001:IA:580470} are used to augment the field lists
so that the location of the appropriate offset can be quickly made.
This is especially beneficial for least-significant fields such as
minutes and seconds.

\begin{figure}[h]
\centering
\includegraphics[width=0.7\columnwidth]{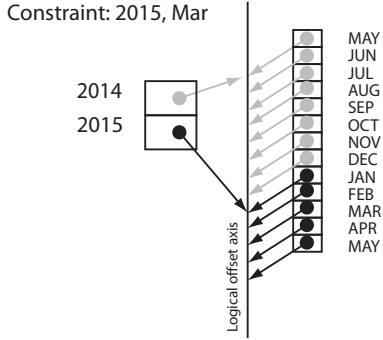}
\caption{Metadata Offset. }
\label{fig:rsmetaoffset}
\end{figure}

Query constraints are applied starting from the most-significant field
(e.g., Year = 2015).  Once the offset is located for the appropriate
year (more than one year may exist), it is used to quickly narrow on
the next constraint (e.g., Month = Mar). That is, the offset for the
matching 2015 year is subsequently used in matching on the interval
tree to find the first matching month within 2015 (see
Figure~\ref{fig:rsmetaoffset}).  Sub-second constraints are realized through
an exponential search (i.e., binary chop).

A persistent version of the in-memory metadata is maintained for fast
recovery. The memory footprint for this index is typically in the
order of tens of megabytes.

\newpage
\subsection{Retrieval APIs}

The LTSS system separates ingress processing and data retrieval across
separate processes.  Data access is read only; the system does not
support insertion or deletion.  Records are implicitly
deleted/overwritten when storage capacity is exhausted (see
Section~\ref{sec:features} on rolling-around).

Queries can be made programmatically through the {\ttcode Record}
{\ttcode Reader} interface, which provides a standard iterator pattern
C++ API. Data can also be accessed through SQL via the SQLite3 virtual
table mechanism~\cite{sqlite3vt}.  The virtual table implementation
backs onto the {\ttcode IRecordReader} interface.  This presents the
data through a table abstraction in which SQL expressions can be
applied.






\section{Implementation Details}
\label{sec:arch}

\subsection{Component Model}
The fundamental building blocks of data pipelines are
{\it actors} and {\it components}.  Actors are memory-protected
processes that exchange control messages through shared memory.
They have at least one thread and typically create new threads for
each client connection.  Messages, based on Google
Protobuf~\cite{protobuf}, are typed and exchanged by either local IPC
or network RPC.  Furthermore, actor messages are either synchronous or
asynchronous (the latter being typical of a stricter actor model).
Actors are composed of components, which are in-process loadable
entities that provide typed interfaces and receptacles that can be
dynamically inspected and bound (much like Microsoft's COM
architecture~\cite{rogerson1997}).  Components do not impose any
threading model.

\subsection{Zero-Copy Data Plane}
Data belonging to flows are exchanged through asynchronous lock-free
FIFO queues.  Units of data, which we term {\it records}, are
exchanged between processes in shared memory and are thus zero-copy.
Record memory is allocated (producer side) and freed (consumer side)
from a thread-safe slab allocator that uses a circular lock-free queue
to track slab elements.  Both the slab allocator queue and the
exchange queue can be configured with SPSC (Single-Producer, Single
Consumer) or MPMC (Multi-Producer, Multi-Consumer) queues depending on
the requirements of the deployment scenario.

\begin{figure}
\centering
\includegraphics[width=0.9\columnwidth]{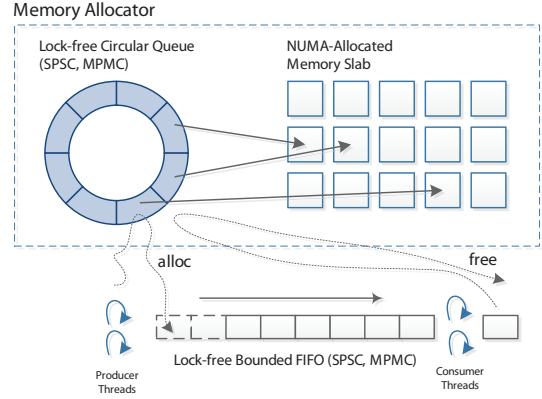}
\caption{Zero-copy IPC}
\label{fig:arch_zipc}
\end{figure}

\newpage

\subsection{NIC Actor}

The prototype LTSS system operates on UDP/IP (version 4) data packets
and makes no assumption about aggregation.  IP provides global
connectivity while UDP provides a simple protocol for
``fire-and-forget'' datagram transmission.  The current prototype does
not use packet authentication or encryption even though this would be
likely in a real-world deployment.  In our experimentation, our sensor
packets are small ($<$ 512 bytes).

\subsection{User-Level Network Stack}

\begin{figure}[h]
\centering
\includegraphics[width=0.9\columnwidth]{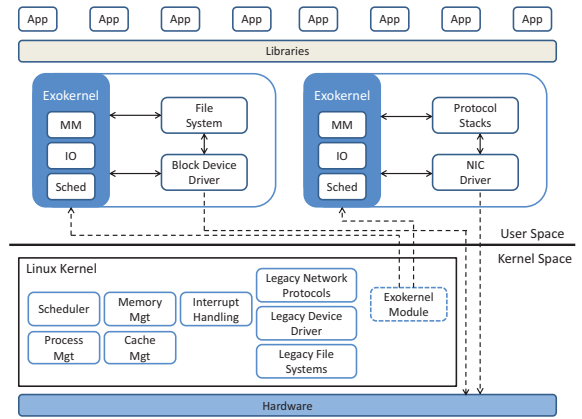}
\caption{Exokernel-based IO Architecture}
\label{fig:exo}
\end{figure}

To maximize network performance, the server-based implementation
adopts an {\it exokernel} design (see Figure~\ref{fig:exo}) that
allows the Network Interface Card (NIC) device driver to fully operate
in user-space~\cite{exokernel}.  In addition to the NIC driver, a
lightweight UDP/IP stack also resides in user-space.  The advantage of
this is improved performance through the elimination of memory copies
(both data and control requests) between the kernel and the
application.  In our experiments, the exokernel approach provides up
to three times the performance of the stock kernel network stack.  We
have not tried an exokernel-based network stack on the embedded
platform because we do not have a user-level device driver for
non-Intel hardware.

\begin{figure}
\centering
\includegraphics[width=1.0\columnwidth]{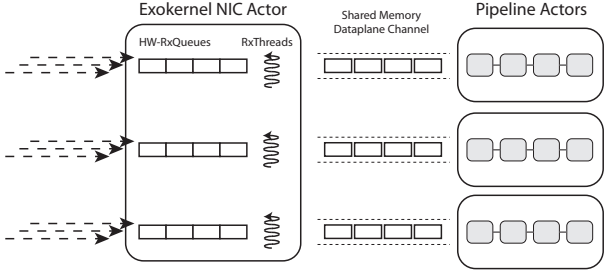}
\caption{Network Data Partitioning Architecture}
\label{fig:arch_hl}
\end{figure}

Each NIC device is managed by a single actor.  The NIC
actor uses multiple Rx threads to service ingress packet processing
for each of the device's hardware queues (see Figure~\ref{fig:arch_hl}).
Load-balancing across queues is achieved through the card's {\it flow
  director} capability, which uses a selective byte in the packet frame to
determine which queue to send the packet to. Each hardware Rx queue also
has its own MSI-X interrupt that is routed to a specific core in which
the Rx thread is also mapped.

Separate data plane channels are established for each Rx thread in the
NIC actor (see Figure~\ref{fig:arch_hl}).  The consumer-side is a
pipeline actor that incorporates processing elements for sorting,
indexing and storing data.  This architecture effectively partitions
data across multiple pipelines enabling scale-up on multicore systems.

LTSS is portable across platforms.  Prototype implementations exist for
both Intel-based commodity x86 servers and workstations, as well as
embedded ARM platforms (e.g., NVIDIA Jetson TK1, Raspberry Pi).  The code
is primarily written in C++.

\subsection{Persistent Store Metadata}  

Record data is written either to a file or directly to a block device
(without any file system).  The latter is useful for integration with
a user-level block device driver such as supported by the exokernel.
In addition to the record data, the Record Store maintains persistent
``block metadata'' that holds information about block usage in the
storage system, partitioning of resources, record count, record size,
etc.  The block metadata is written in 4K blocks.  $N$ copies of the
block metadata are stored on disk, where $N$ is the ``rolling count''.
Three is typical.  When the block device does not support atomic 4K
writes, a checksum is used to verify integrity.


\subsection{SQLite3 Integration}

\begin{figure}
\centering
\includegraphics[width=0.98\columnwidth]{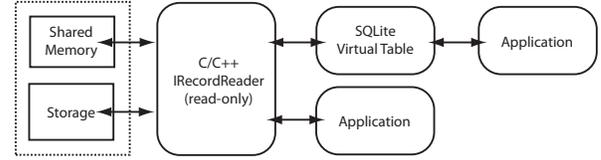}
\caption{Read-only Query Side.}
\label{fig:sqlinteg}
\end{figure}

The LTSS system supports both programmatic and SQL-based query
interfaces (see Figure~\ref{fig:sqlinteg}).  SQL support is realized
through the virtual table mechanism~\cite{sqlite3vt}, which uses a
custom loadable library to tailor data storage and retrieval below the
SQL query engine.  The LTSS virtual table implementation is limited to
read-only queries.

The LTSS virtual table implements methods for connection management
({\ttcode xCreate},{\ttcode xConnect}, {\ttcode xDisconnect},{\ttcode xDestroy}), query plan
exploration ({\ttcode xBestIndex}), constraint filtering ({\ttcode
  xFilter}), and iterator-based data retrieval ({\ttcode xNext, xEof,
  xColumn, xRowid}). The underlying record store supports constraints
on one or more composite time fields and any other numeric field.
Other constraints are delegated to the SQLite3 query engine, which can
perform filtering above the virtual table implementation.

\subsection{Parallel Queries}

LTSS supports parallel SQL queries across multiple data partitions
(i.e. pipelines).  Re-ordering constraints (e.g., {\ttcode ORDER BY})
are handled by the SQL engine.  The current implementation supports
combining results through append operations (i.e., the iterator first
exhausts partition $A$, then partition $B$ and so forth) and sort-merge operations.

\subsection{Other Features}
\label{sec:features}

{\it Pre-fetching} - The system optionally supports using main memory
to cache and pre-fetch query results.  This can be advantageous to
lower selectivity queries.  The amount of main memory allocated for
pre-fetching is fully configurable.

{\it Storage Roll-around} - Because of the simplistic design around
record storage, the LTSS can easily ``roll-around'' the storage so
that the oldest records are overwritten by new records according to
some total space allocation.  The LTSS system supports both file-based
or direct (block-level) storage.  Direct storage supports allocation
of block zones in order to sub-divide the raw block capacity.

{\it Continuous Queries} - Current support for continuous queries is
limited.  The implementation supports the use of callbacks on updates
so that the client can be notified when a given number of new records
have been inserted.  This can be used to trigger summarization or
aggregation functions (e.g., over a sliding window) that can be
accumulated in a separate table.

\section{Evaluation} 
\label{sec:eval}

To evaluate the performance of the LTSS we used three publicly
available data sets; {\it Seismic}, {\it Taxi} and {\it Energy} (see
Table~\ref{tbl:data_src}).

\begin{table}[ht]
\begin{center}
\begin{tabularx}{1.0\linewidth}{|
>{\setlength{\hsize}{.15\hsize}\raggedright\scriptsize}X|
>{\setlength{\hsize}{.55\hsize}\raggedright\scriptsize}X|
>{\setlength{\hsize}{.13\hsize}\raggedright\scriptsize}X|
>{\setlength{\hsize}{.17\hsize}\raggedright\arraybackslash\scriptsize}X|}\hline

{\bf Name} & {\bf Description \& Source of Data}  & {\bf Record Size (bytes)} & {\bf Sample Size (records)} \\\hline

Seismic & USGS archived earthquake data (http://earthquake.usgs.gov) & 28 & 2.810 B \\\hline
Taxi    & NYC taxi data from Andrés Monroy (http://chriswhong.com/ open-data/foil\_nyc\_taxi/) & 132 & 169.9 M \\\hline 
Energy  & Household energy use from Belkin Energy Disaggregation Kaggle competition (https://www.kaggle.com/c/ belkin-energy-disaggregation-competition) & 119 & 14.7 M \\\hline
 
\end{tabularx}
\end{center}
\caption{Experimental Data Sources used for Evaluation Experiments.}
\label{tbl:data_src}
\end{table}

\begin{figure}
\centering
\includegraphics[width=0.8\columnwidth]{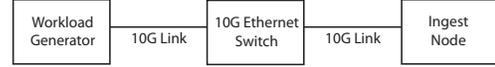}
\caption{Experimental Setup for Ingestion Measurements.}
\label{fig:exp_ing}
\end{figure}

Use of public data, as opposed to proprietary data,
enables others to make a fair comparison.  Data is taken from archives
of actual data and replayed with high-fidelity using our time-series
workload generator~\cite{Kuang2015}.  For ``stress'' load generation
(i.e. faster than the original source rate) the replay is artificially
re-stamped.  Data is transmitted from a dedicated generator server
across a 10G switched link (see Figure~\ref{fig:exp_ing}). Each data
sample is encapsulated in a single IP/UDP packet (i.e., no aggregation
occurs).

Seven representative benchmark queries were constructed for each of
the Energy and Taxi data sets.  These benchmarks were specifically
designed to measure the performance of queries where time-based
columns are the primary query constraints.  We purposely avoided
existing SQL benchmark suites (e.g., TPC) that test a range of queries
beyond the scope of the LTSS design intent.  The SQL query code is
given in the appendix.  correctness.  The Seismic data set was used
for sliding-window queries for the purpose of read-write contention
measurement.

It is not practical to compare the performance of LTSS with all of the
previously discussed time-series database solutions (refer to Section
2).  To at least provide a baseline we compare the performance of LTSS
with SQLite3 and PostgreSQL 9.4.  We chose SQLite3 because this is the
basis for our own solution and it can also be deployed in embedded
environments.  We chose PostgreSQL because this is the basis of
Pipeline DB; an open source time-series data with comparable features.
Results were cross-verified on each to ensure correctness.

\subsection{System Configuration}

For evaluation, the LTSS was deployed on a unloaded systems.  Both a
commodity x86 server and embedded ARM (Raspberry Pi) platform.
Details of the HW are given in Table~\ref{tbl:platform_spec}.

\begin{table}[H]
\begin{center}
\begin{tabularx}{1.0\linewidth}{|
>{\setlength{\hsize}{.3\hsize}\raggedright\scriptsize}X|
>{\setlength{\hsize}{.7\hsize}\raggedright\arraybackslash\scriptsize}X|}\hline

\multirow{4}{20mm}{\bf{Commodity x86 Server}} &
Dell R720 Server with Intel E5-2670 v2 @ 2.5GHz CPU  \\\cline{2-2} &  
10 cores, 20 hardware threads per socket \\\cline{2-2} & 
32GB DRAM, 25MB shared L3 cache, per-core 256KB L2 cache and 32KB L1 cache \\\cline{2-2} &  
Intel X540 10 Gbps NIC (x8 PCIe v2.1, 5GT/s) \\\hline

\hline

\multirow{4}{20mm}{\bf{Embedded Platform}} &
Raspberry Pi 2 Model B+, with Broadcom BCM2836 ARM Cortex-A7 @ 900 MHz\\\cline{2-2} &  
4 cores, 1GB DRAM \\\cline{2-2} & 
64GB external Transcend 30MB/s SD Card \\\cline{2-2} & 
On-board 10/100MBps Ethernet \\\hline

\end{tabularx}
\end{center}
\caption{Test Platform Specifications.}
\label{tbl:platform_spec}
\end{table}

\renewcommand{\multirowsetup}{\centering}

\newpage
\subsection{Ingestion Performance}

The first performance measure is ingestion rate in terms of {\it maximum
  throughput}.  Maximum throughput is defined as the maximum ingestion
rate with zero-packet/record loss and without ``delinquent''
packets.  A packet is considered delinquent when it arrives at the
system with a time stamp corresponding to a quantum bucket that has
already been closed (i.e., the respective time window has already
been passed to the Record Store).

To measure the equivalent network-based ingestion performance of
SQLite3 and PostgreSQL, we built wrapper components that can be
connected directly to the ChannelSource (refer to
Figure~\ref{fig:arch_tihlv}).  Exceeding maximum throughput results in
packets being dropped by the NIC-actor (due to blocking on the
down-stream components).  Measurements were taken for a single
pipeline (using wrappers for SQLite3 and PostgreSQL).  For PostgreSQL,
the records where written with the {\ttcode COPY} command which
inserts records faster than the {\ttcode INSERT} command.  For SQLite,
we do not use Write-Ahead Logging (WAL) mode since this will lock out read-access
to the database.  To perform insertions we used {\ttcode INSERT}s batched
using {\ttcode BEGIN},{\ttcode END} transaction primitives.

Table~\ref{tbl:ingestcomp} shows a comparison of ingestion rates for
the three systems on the server
platform. Table~\ref{tbl:ingestcompembed} shows ingestion rate for the
Seismic data on LTSS and SQLite3.  The results show ingestion rates
and {\it load index} (in brackets).  Load index corresponds to the
mean sum of CPU load.  For example, a load of 2.0 is equivalent to two
logical cores 100\% active or four cores 50\% all of the time. We did
not measure the performance of PostgreSQL on the embedded platform
since it was not designed with resource constrained environments in
mind.

\setlength{\extrarowheight}{3pt}

\begin{table}
{\scriptsize
\begin{tabularx}{1.0\linewidth}{|
 >{\setlength{\hsize}{.25\hsize}\raggedright\scriptsize}X|
 >{\setlength{\hsize}{.25\hsize}\raggedright\scriptsize}X|
 >{\setlength{\hsize}{.25\hsize}\raggedright\scriptsize}X|
 >{\setlength{\hsize}{.25\hsize}\raggedright\arraybackslash\scriptsize}X|}\hline

\multirow{2}{*}{\bf Data Set} & \multicolumn{3}{c|}{\bf Maximum Ingest Throughput (Load Index)}\\\cline{2-4}
& {\bf SQLite3} & {\bf PostgreSQL} & {\bf LTSS}\\\hline
Seismic &  332K (1.85)  &  299K (1.38) &  1142K (2.80) \\\hline
Taxi    &  242K (2.20)  &  267K (1.75) &   850K (2.42) \\\hline
Energy  &  220K (1.81)  &  203K (1.59) &   913K (2.80) \\\hline

\end{tabularx}
\caption{Single Pipeline Throughput (Server).}\label{tblfp}
\label{tbl:ingestcomp}
}
\end{table}

\begin{table}
{\scriptsize
\begin{tabularx}{1.0\linewidth}{|
 >{\setlength{\hsize}{.2\hsize}\raggedright\scriptsize}X|
 >{\setlength{\hsize}{.4\hsize}\raggedright\scriptsize}X|
 >{\setlength{\hsize}{.4\hsize}\raggedright\arraybackslash\scriptsize}X|}\hline

\multirow{2}{*}{\bf Data Set} & \multicolumn{2}{c|}{\bf Maximum Ingest Throughput (Load Index)}\\\cline{2-3}
& {\bf SQLite3} & {\bf LTSS}\\\hline
Seismic &  25K (1.97)  & 69K (2.59) \\\hline

\end{tabularx}
\caption{Single Pipeline Throughput (Embedded).}\label{tblfp}
\label{tbl:ingestcompembed}
}
\end{table}

\setlength{\extrarowheight}{0pt}

The data shows that LTSS delivers a factor of between 2x and 4x
increase in ingestion throughput over the other systems and a rate-per-core
improvement factor (calculated by the load index) of between 2x and 3x.

\subsubsection{Ingestion Scaling}

We also explored the scaling of LTSS ingestion rate with an increasing
number of pipelines (and hence data partitions).  To do this, we
measured ingestion performance of Seismic data up to the maximum
sustainable (6 pipelines) by a single NIC card.  We chose Seismic data for
this test because its small size would best stress the IO handling.

\begin{figure}[h]
\centering
\includegraphics[width=0.8\columnwidth]{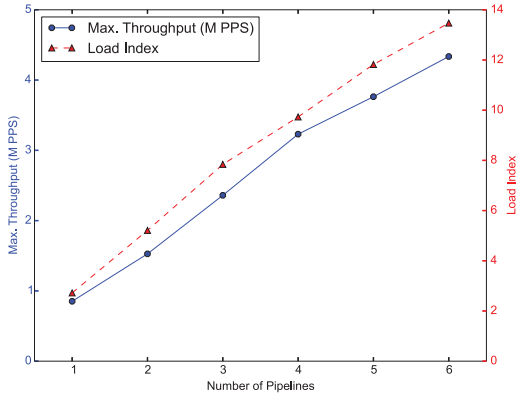}
\caption{LTSS Ingestion Scaling (multi-pipeline).}
\label{fig:ingestscale}
\end{figure}

We observed a maximum throughput of 4.33 M records per second with 6
data pipelines (one record per packet) and a load index of 13.47.
This corresponds to approximately 34\% system CPU load.  We could not
increase the number of pipelines without jeopardizing jitter and
consequently resulting in packet loss.  Figure~\ref{fig:ingestscale}
shows the ingestion scaling for LTSS.  Note that the y-axis shows
both maximum throughput and load index.

\subsection{Storage Footprint}

We measured the on-disk storage footprint (of a clean database)
for each of the three data sets.  The data schema was the same across
each database except that the LTSS exposes multiple logical fields for
composite time.

\setlength{\extrarowheight}{3pt}
\begin{table}[!h]
{\scriptsize
\begin{tabularx}{0.97\linewidth}{|
 >{\setlength{\hsize}{.17\hsize}\raggedright\scriptsize}X|
 >{\setlength{\hsize}{.17\hsize}\raggedright\scriptsize}X|
 >{\setlength{\hsize}{.21\hsize}\raggedright\scriptsize}X|
 >{\setlength{\hsize}{.24\hsize}\raggedright\scriptsize}X|
 >{\setlength{\hsize}{.18\hsize}\raggedright\arraybackslash\scriptsize}X|}\hline

\multirow{2}{*}{\bf Data Set} & \multirow{2}{*}{\bf Records} & \multicolumn{3}{c|}{\bf Storage Footprint (MB)}\\\cline{3-5}
& & {\bf SQLite3} & {\bf PostgreSQL} & {\bf LTSS}\\\hline
Seismic &  5.43B   &  179,913  &  246,937  &  186,436 \\\hline
Taxi    &  169.6M  &  29,926   &  32,161   &  23,760 \\\hline
Energy  &  14.7M   &  3,815    &  5,874    &  1,761 \\\hline

\end{tabularx}
\caption{Storage Footprints.}\label{tblfp}
}
\end{table}
\setlength{\extrarowheight}{0pt}

The storage footprints for each data set are given in
Table~\ref{tblfp}. Note, data represents the footprint before
queries are performed (i.e., before any storage is used for query
caching).  The data indicates that LTSS in many cases has a smaller
storage footprint than the other two solutions.  Of course,
compression could also be integrated into LTSS should footprint
be of concern.

\subsection{Memory Footprint}

Run-time memory footprint is another important factor to consider.  We
measured memory footprints for both server and embedded deployments
after ingesting 100M records. In all case we exclude the overhead of
the NIC actor and its buffer allocation ($\sim$18 MB).  We used the default
configurations for SQLite3 and PostgreSQL, and buffered records into
4MB chunks for each transaction.  For PostgreSQL, we exclude overhead
(in the order of hundreds of MB) caused by related processes for check
pointing, stats collection, etc.  The aim of these measurements is to
get an understanding of {\it typical} run-time overheads.  SQLite3 and 
PostgreSQL may be configurable with less memory but that likely will
change ingestion and query performance.

\setlength{\extrarowheight}{3pt}
\begin{table}[!h]
{\scriptsize
\begin{tabularx}{1.0\linewidth}{|
 >{\setlength{\hsize}{.2\hsize}\raggedright\scriptsize}X|
 >{\setlength{\hsize}{.18\hsize}\raggedright\scriptsize}X|
 >{\setlength{\hsize}{.24\hsize}\raggedright\scriptsize}X|
 >{\setlength{\hsize}{.18\hsize}\raggedright\scriptsize}X|
 >{\setlength{\hsize}{.2\hsize}\raggedright\arraybackslash\scriptsize}X|}\hline

{\bf SQLite3 (server)} & {\bf SQLite3 (embedded)} & {\bf PostgreSQL (server)} & {\bf LTSS (server)} & {\bf LTSS (embedded)} \\\hline
$\sim$323 MB                 & 35 MB                    & $\sim$335 MB                    & 27 MB               & 16 MB \\\hline


\end{tabularx}
\caption{Memory Footprints.}\label{tblfp}
}
\end{table}
\setlength{\extrarowheight}{0pt}

\newpage
\subsection{Out-of-order Tolerance}

To examine the tolerance for out-of-order (generally late) packets, we
artificially re-stamped packets.  We measured the effect on maximum
performance for a single pipeline ingesting the Seismic data set.  Two
modes were measured, adding a fixed delay of 80ms, and adding a random
delay of between 1 and 100 ms.  Delays were added according to fixed
ratios (1:100,1:10,1:5 and 1:2).  Table~\ref{tbl:oootol} shows the results.

\begin{table}[h]
\begin{center}
\begin{tabularx}{0.8\linewidth}{|
>{\setlength{\hsize}{.2\hsize}\raggedright\scriptsize}X|
>{\setlength{\hsize}{.3\hsize}\raggedright\scriptsize}X|
>{\setlength{\hsize}{.3\hsize}\raggedright\arraybackslash\scriptsize}X|}\hline

\multirow{2}{*}{\bf Ratio} & \multicolumn{2}{c|}{\scriptsize\bf Max. Throughput (K PPS)}\\\cline{2-3}
& {\bf 80ms fixed} & {\bf 100ms random} \\\hline

1:100 & 844.8 & 981.5 \\\hline
1:10 & 751.0 & 713.6 \\\hline
1:5 & 748.4 & 672.1 \\\hline
1:2 & 734.9 & 587.6 \\\hline

\end{tabularx}
\end{center}
\caption{LTSS Out-of-order Tolerance for Seismic Data.}
\label{tbl:oootol}
\end{table}

The data indicates that LTSS is reasonably tolerant of out-of-order
and would adequately handle expected out-of-order data with little
performance degradation.

\subsection{Query (Read) Performance}

Next, we evaluate SQL query performance.  Measurements were taken over
50 consecutive runs on a unloaded system without active queries.  The
databases were stored on NVMe SSD media via an ext4 file system.  The
SQLite3 and PostgreSQL database instances include a secondary index
for the time column (i.e. using the {\ttcode CREATE INDEX} SQL
construct).  Note that the SQLite3 and PostgreSQL do not support any
form of {\it patterned} time type and therefore conversion operators
need (e.g., {\ttcode EXTRACT} and {\ttcode strftime}) to be applied at
query time.  Also, in LTSS, the composite time is derived at ingest 
time, not at query time.

\begin{figure}[h]
\centering
\includegraphics[width=0.98\columnwidth]{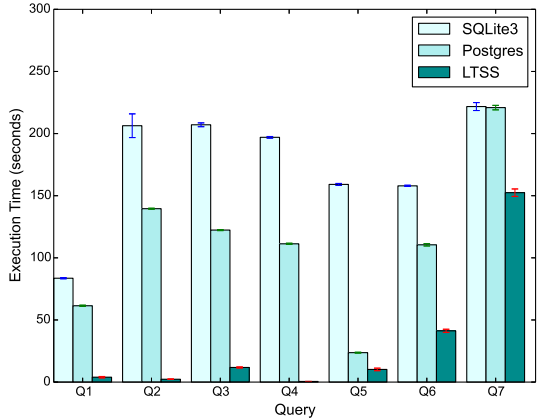}
\caption{NYC Taxi Query Performance.}
\label{fig:taxiplot}
\end{figure}

\begin{figure}[h]
\centering
\includegraphics[width=0.98\columnwidth]{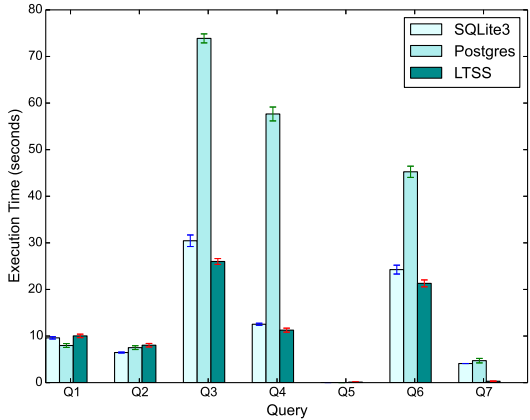}
\caption{Belkin Energy Query Performance.}
\label{fig:belkinplot}
\end{figure}

The results shown in Figure~\ref{fig:taxiplot} and Figure~\ref{fig:belkinplot}
compare query performance across the three systems for the Taxi and
Energy benchmarks (see Appendix for query detail).  We focus on these
two data sets for evaluation because their queries are more complex and
incorporate multiple attributes (e.g., sliding windows).

LTSS is consistently better in terms of query performance and provides
significant improvements where the time dimension is highly selective
(e.g., Q4 Taxi).

\subsection{Query (Read) Performance under Ingestion Load}

Finally, we measured the effect of ingest load on query performance
(i.e., read-write contention) using the Seismic data set. We chose
this data set because of its small record size and thus high packet
rate.  The test was carried out by applying an aggregation operator
({\it average}) over a sliding window query of the last 1000 inserted
records.  This is given with the following query:

{\scriptsize
\begin{verbatim}
  WITH vals(v) AS 
    (SELECT value FROM seismic LIMIT 1000) 
  SELECT avg(v) FROM vals;
\end{verbatim}
}

\begin{figure}
\centering
\includegraphics[width=0.98\columnwidth]{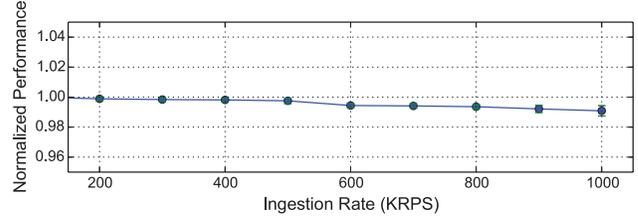}
\caption{Read-Write Contention for LTSS.}
\label{fig:wrcontent}
\end{figure}

Query rate (QPS) was taken as a mean across 40 runs each with 20
windows.  The data in Figure~\ref{fig:wrcontent} shows fluctuations in
performance of less that 1\% on increasing ingestion rate up to 1M
events/sec in a single pipeline.

We do not include results for read-write contention on SQLite3 and PostgreSQL
because they perform badly due to their locking architectures.  We
were unable to avoid packet drop under any reasonable query load (even 
with the client opened in read-only mode).

\section{Conclusions} \label{sec:arch}

Continued evolution of IoT will bring with it a need to aggregate,
process and store large volumes of time-series data.  In this paper we
have presented a lightweight architecture for storage and indexing of
time-series data streams within the context of IoT.  Our system, LTSS,
is based on a software pipeline architecture that is optimized for
multicore processing and parallel IO capabilities.  The architecture
is specifically designed to take advantage of fast, random-access
friendly, non-volatile storage devices (e.g., NVMe) and aims to reduce
the dependency on DRAM by simplifying in-memory data structures.  The
result is a design that can be readily deployed on both commodity
server as well as resource constrained and embedded environments.

Our results show a data ingest (completely to persistent store) in the
order of 4 million RPS (without UDP packing) for a single 10G NIC with
a load index of 14.  For the Raspberry Pi 2 embedded platform, we show
an ingest capability of around 70K RPS.  These results are 3x to 4x
improvement over existing RDBMS solutions.  We also show that ad-hoc
query performance, using SQL, is significantly better than that
achievable with two prominent databases, SQLite3 and Postgres.  Our
solution currently does not employ compression although there is no
fundamental reason why this could not be added should a need arise.

We believe that this work represents a first step to considering the
increasing shift from IO-bound to CPU-bound data processing and
management.  Careful consideration of data flow across multiple
processor cores and optimized memory management is now becoming a
fundamental element of successful high-performance stream processing
and storage designs.





\bibliographystyle{abbrv} 
\bibliography{references}

\begin{thebibliography}{10}

\bibitem{exokernel}
Exokernel development kit (xdk).
\newblock \url{https://github.com/dwaddington/xdk}.
\newblock Accessed: 2016-01-06.

\bibitem{informix}
Ibm informix timeseries data user's guide.
\newblock Accessed: 2016-01-21.

\bibitem{infiniflux}
Infiniflux: The world's fastest time series dbms for iot and big data.
\newblock \url{http://www.infiniflux.com/}.
\newblock Accessed: 2016-01-21.

\bibitem{parstream}
Parstream db: The engine inside parstream’s analytics platform for iot.
\newblock \url{https://www.parstream.com/product/parstream-db/}.
\newblock Accessed: 2016-01-14.

\bibitem{pipelinedb}
Pipeline db documentation.
\newblock \url{http://docs.pipelinedb.com/}.
\newblock Accessed: 2016-01-21.

\bibitem{protobuf}
Protocol buffers.
\newblock \url{https://developers.google.com/protocol-buffers/}.
\newblock Accessed: 2016-01-04.

\bibitem{sap}
Sap businessobject business intelligence.
\newblock \url{http://go.sap.com/product/analytics/bi-platform.html}.
\newblock Accessed: 2016-01-26.

\bibitem{sas}
Sas analytics platform.
\newblock \url{http://support.sas.com/documentation/onlinedoc/apcore/}.
\newblock Accessed: 2016-01-26.

\bibitem{tableau}
Tableau business intelligence.
\newblock \url{http://www.tableau.com/business-intelligence}.
\newblock Accessed: 2016-01-26.

\bibitem{parstreamwp}
Technical whitepaper: Real-time database for big data analytics.
\newblock \url{http://static1.squarespace.com/}.
\newblock Accessed: 2016-01-14.

\bibitem{sqlite3vt}
The virtual table mechanism of sqlite.
\newblock \url{https://www.sqlite.org/vtab.html}.
\newblock Accessed: 2015-12-10.

\bibitem{Cook:1980:BSA:359024.359026}
C.~R. Cook and D.~J. Kim.
\newblock Best sorting algorithm for nearly sorted lists.
\newblock {\em Commun. ACM}, 23(11):620--624, Nov. 1980.

\bibitem{Cormen:2001:IA:580470}
T.~H. Cormen, C.~Stein, R.~L. Rivest, and C.~E. Leiserson.
\newblock {\em Introduction to Algorithms}.
\newblock McGraw-Hill Higher Education, 2nd edition, 2001.

\bibitem{Elmasri:1999:FDS:554501}
R.~A. Elmasri and S.~B. Navathe.
\newblock {\em Fundamentals of Database Systems}.
\newblock Addison-Wesley Longman Publishing Co., Inc., Boston, MA, USA, 3rd
  edition, 1999.

\bibitem{Groff:2009:SCR:1594046}
J.~Groff and P.~Weinberg.
\newblock {\em SQL The Complete Reference, 3rd Edition}.
\newblock McGraw-Hill, Inc., New York, NY, USA, 3 edition, 2010.

\bibitem{parstream:patent}
N.~H. Jorg~Bienert, Michael~Hummel.
\newblock Method and system for compressing data records and for processing,
  August 2013.

\bibitem{Kuang2015}
J.~Kuang, D.~G. Waddington, and C.~Lin.
\newblock Techniques for fast and scalable time series traffic generation.
\newblock In {\em Proceedings of the 2015 IEEE Conference on Big Data},
  BigData'15, Oct. 2015.

\bibitem{O'Neil96thelog-structured}
P.~O'Neil, E.~Cheng, D.~Gawlick, and E.~O'Neil.
\newblock The log-structured merge-tree (lsm-tree), 1996.

\bibitem{rogerson1997}
D.~Rogerson.
\newblock {\em COM}.
\newblock Microsoft programming series. Microsoft Press, 1997.

\bibitem{Zaharia:2010:SCC:1863103.1863113}
M.~Zaharia, M.~Chowdhury, M.~J. Franklin, S.~Shenker, and I.~Stoica.
\newblock Spark: Cluster computing with working sets.
\newblock In {\em Proceedings of the 2Nd USENIX Conference on Hot Topics in
  Cloud Computing}, HotCloud'10, pages 10--10, Berkeley, CA, USA, 2010. USENIX
  Association.

\end{thebibliography}

\onecolumn

\begin{appendix}

\section{SQLite3 Taxi Data Benchmark Queries}
\label{sec:sqlitetaxiqueries}
{\scriptsize
\begin{verbatim}
Q1. Count number of pickups after 8pm. 
SELECT count(*) FROM TAXIS
WHERE CAST(strftime('%H',pickup_datetime, 'unixepoch', 'localtime') AS INTEGER) >= 20;
---
Q2. Count number of weekday picks in November 2013.
SELECT count(*) FROM TAXIS
WHERE CAST(strftime('%w',pickup_datetime, 'unixepoch', 'localtime') AS INTEGER) > 0
AND CAST(strftime('%w',pickup_datetime, 'unixepoch', 'localtime') AS INTEGER) < 6
AND CAST(strftime('%m',pickup_datetime, 'unixepoch', 'localtime') AS INTEGER) = 11
AND CAST(strftime('%Y',pickup_datetime, 'unixepoch', 'localtime') AS INTEGER) = 2013;
---
Q3. Average time of weekday trips in Summer (Jun-Oct).
SELECT avg(trip_time_in_secs) FROM TAXIS
WHERE CAST(strftime('%m', pickup_datetime, 'unixepoch', 'localtime') AS INTEGER) >= 6
AND CAST(strftime('%m', pickup_datetime, 'unixepoch', 'localtime') AS INTEGER) <= 10
AND CAST(strftime('%w', pickup_datetime, 'unixepoch', 'localtime') AS INTEGER) > 0
AND CAST(strftime('%w', pickup_datetime, 'unixepoch', 'localtime') AS INTEGER) < 6;
---
Q4. Shortest and longest trips on the day of 11/25/2013.
SELECT min(trip_time_in_secs), max(trip_time_in_secs) FROM TAXIS
WHERE CAST(strftime('%Y',pickup_datetime, 'unixepoch', 'localtime') AS INTEGER) = 2013
AND CAST(strftime('%m',pickup_datetime, 'unixepoch', 'localtime') AS INTEGER) = 11
AND CAST(strftime('%d',pickup_datetime, 'unixepoch', 'localtime') AS INTEGER) = 25;
---
Q5. Total trip distance for a specific vehicle between 9am and 12 noon.
SELECT sum(trip_distance) FROM TAXIS
WHERE CAST(strftime('%H',pickup_datetime, 'unixepoch', 'localtime') AS INTEGER) >= 9
AND CAST(strftime('%H',pickup_datetime, 'unixepoch', 'localtime') AS INTEGER) < 12
AND medallion = '5CC9B3C9725FCD7FAE490B4C614D57EE';
---
Q6. Total number of passengers on Saturday and Sunday.
SELECT sum(passenger_count) FROM TAXIS
WHERE CAST(strftime('%w',pickup_datetime, 'unixepoch', 'localtime') AS INTEGER) == 0
OR CAST(strftime('%w',pickup_datetime, 'unixepoch', 'localtime') AS INTEGER) == 6;
---
Q7. Total number of passengers each day of the week.
SELECT CAST(strftime('%w',pickup_datetime, 'unixepoch', 'localtime') AS INTEGER),sum(passenger_count) FROM TAXIS
GROUP BY CAST(strftime('%w',pickup_datetime, 'unixepoch', 'localtime') AS INTEGER);
\end{verbatim}
}

\section{Postgres Taxi Data Benchmark Queries}
\label{sec:ptaxiqueries}
{\scriptsize
\begin{verbatim}
Q1. Count number of pickups after 8pm. 
SELECT count(*) FROM TAXI WHERE EXTRACT(hour FROM 
pickup_datetime::int4::abstime::timestamp) >= 20;
---
Q2. Count number of weekday picks in November 2013.
SELECT count(*) FROM TAXI
WHERE EXTRACT(dow FROM pickup_datetime::int4::abstime::timestamp) > 0 
AND EXTRACT(dow FROM pickup_datetime::int4::abstime::timestamp) < 6
AND EXTRACT(month FROM pickup_datetime::int4::abstime::timestamp) = 11
AND EXTRACT(year FROM pickup_datetime::int4::abstime::timestamp) = 2013;
---
Q3. Average time of weekday trips in Summer (Jun-Oct).
SELECT avg(trip_time_in_secs) FROM TAXI
WHERE EXTRACT(month FROM pickup_datetime::int4::abstime::timestamp) >= 6
AND EXTRACT(month FROM pickup_datetime::int4::abstime::timestamp) <= 10
AND EXTRACT(dow FROM pickup_datetime::int4::abstime::timestamp) > 0
AND EXTRACT(dow FROM pickup_datetime::int4::abstime::timestamp) < 6;
---
Q4. Shortest and longest trips on the day of 11/25/2013. 
SELECT min(trip_time_in_secs), max(trip_time_in_secs) FROM TAXI
WHERE EXTRACT(year FROM pickup_datetime::int4::abstime::timestamp) = 2013
AND EXTRACT(month FROM pickup_datetime::int4::abstime::timestamp) = 11
AND EXTRACT(day FROM pickup_datetime::int4::abstime::timestamp) = 25;
---
Q5. Total trip distance for a specific vehicle between 9am and 12 noon.
SELECT sum(trip_distance) FROM TAXI
WHERE EXTRACT(hour FROM pickup_datetime::int4::abstime::timestamp) >= 9
AND EXTRACT(hour FROM pickup_datetime::int4::abstime::timestamp) < 12
AND medallion = '5CC9B3C9725FCD7FAE490B4C614D57EE';
---
Q6. Total number of passengers on Saturday and Sunday.
SELECT sum(passenger_count) FROM TAXI
WHERE EXTRACT(dow FROM pickup_datetime::int4::abstime::timestamp) = 0
OR EXTRACT(dow FROM pickup_datetime::int4::abstime::timestamp) = 6;
---
Q7. Total number of passengers each day of the week.
SELECT EXTRACT(dow FROM pickup_datetime::int4::abstime::timestamp), 
sum(passenger_count) FROM TAXI
GROUP BY EXTRACT(dow FROM pickup_datetime::int4::abstime::timestamp);
\end{verbatim}
}

\section{LTSS Taxi Data Benchmark Queries}
\label{sec:tsstaxiqueries}
{\scriptsize
\begin{verbatim}
Q1. Count number of pickups after 8pm. 
SELECT count(*) FROM TAXI WHERE CTIME_pickup_hour >= 20;
---
Q2. Count number of weekday picks in November 2013.
SELECT count(*) FROM TAXI WHERE CTIME_pickup_wday > 0 
AND CTIME_pickup_wday < 6 
AND CTIME_pickup_month = 11 AND CTIME_pickup_year = 13;
---
Q3. Average time of weekday trips in Summer (Jun-Oct).
SELECT avg(trip_time_in_secs) FROM TAXI
WHERE CTIME_pickup_month >= 6 AND CTIME_pickup_month <= 10
AND CTIME_pickup_wday > 0 AND CTIME_pickup_wday < 6;
---
Q4. Shortest and longest trips on the day of 11/25/2013. 
SELECT min(trip_time_in_secs), max(trip_time_in_secs) FROM TAXI
WHERE CTIME_pickup_year = 13 AND CTIME_pickup_month = 11 
AND CTIME_pickup_day = 25;
---
Q5. Total trip distance for a specific vehicle between 9am and 12 noon.
SELECT sum(trip_distance) FROM TAXI WHERE CTIME_pickup_hour >= 9
AND CTIME_pickup_hour < 12 
AND  medallion = '5CC9B3C9725FCD7FAE490B4C614D57EE';
---
Q6. Total number of passengers on Saturday and Sunday.
SELECT sum(passenger_count) FROM TAXI WHERE CTIME_pickup_wday == 0 
OR CTIME_pickup_wday == 6;
---
Q7. Total number of passengers each day of the week.
SELECT CTIME_pickup_wday,sum(passenger_count) FROM TAXI 
GROUP BY CTIME_pickup_wday;
\end{verbatim}
}

\section{SQLite3 Energy Data Benchmark Queries}
\label{sec:sqliteeneryqueries}
{\scriptsize 
\begin{verbatim}Q1. Hourly average power consumption for house H1.
SELECT strftime('%H',DATETIME), avg(V0 * I0) FROM POWER
WHERE HOUSEID = 'H1' GROUP BY strftime('%H',DATETIME);
---
Q2. Maximum power sample for each house.
SELECT HOUSEID, max(V0*I0) FROM POWER
WHERE CAST(strftime('%H',DATETIME) AS INTEGER) > 8 AND CAST(strftime('%H',DATETIME) AS INTEGER) <= 20
GROUP BY HOUSEID ORDER BY HOUSEID;
---
Q3. Highest hourly average power sample point for each house.
WITH 
     hourlies (HOUSEID, HOUR, POWER) AS (SELECT HOUSEID, strftime('%H',DATETIME), avg(V0 * I0) FROM POWER 
GROUP BY HOUSEID, strftime('%H',DATETIME) ORDER BY avg(V0*I0) DESC)
SELECT HOUSEID, HOUR, max(POWER) FROM hourlies
WHERE HOUSEID IN (SELECT DISTINCT HOUSEID FROM POWER)
---
Q4. Top ten, 5 minute periods of consumption from house 'H1' (tumbling window).
SELECT HOUSEID, avg(V0*I0), (strftime('%s',DATETIME)/300) FROM POWER 
WHERE HOUSEID='H1' GROUP BY HOUSEID, (strftime('%s',DATETIME)/300) ORDER BY (strftime('%s',DATETIME)/300) DESC LIMIT 10;
---
Q5. Number of samples between two datetimes
SELECT count(*) FROM POWER WHERE DATETIME >= '2012-07-30 09:35:00' AND DATETIME <= '2012-07-30 09:39:00';
---
Q6. Average maxium weekday consumption for each house.
WITH 
     weekday_max(houseid, wdaymax, power)
     AS (
         SELECT HOUSEID, strftime('%w',DATETIME), max(V0*I0)
         FROM POWER
         GROUP BY HOUSEID, (strftime('%w',DATETIME))
        )
SELECT houseid, avg(power) FROM weekday_max GROUP BY houseid ORDER BY houseid;
---
Q7. Number of samples taken between 5pm and 9pm on Wednesday.
SELECT count(*) FROM POWER WHERE strftime('%w',DATETIME) = '3' 
AND CAST(strftime('%H',DATETIME) as INTEGER) >= 17 AND CAST(strftime('%H',DATETIME) as INTEGER) <= 20;
\end{verbatim}
}

\section{Postgres Energy Data Benchmark Queries}
\label{sec:peneryqueries}
{\scriptsize 
\begin{verbatim}
Q1. Hourly average power consumption for house H1.
SELECT EXTRACT(hour FROM DATETIME), avg(V0 * I0) FROM POWER
WHERE HOUSEID = 'H1' GROUP BY EXTRACT(hour FROM DATETIME);
---
Q2. Maximum power sample for each house.
SELECT HOUSEID, max(V0*I0) FROM POWER
WHERE EXTRACT(hour FROM DATETIME) > 8 AND EXTRACT(hour FROM DATETIME) <= 20
GROUP BY HOUSEID ORDER BY HOUSEID;
---
Q3. Highest hourly average power sample point for each house.
WITH hourlies (HOUSEID, HOUR, POWER) AS (
              SELECT HOUSEID, EXTRACT(hour FROM DATETIME), avg(V0 * I0) 
              FROM POWER 
              GROUP BY HOUSEID, EXTRACT(hour FROM DATETIME) 
              ORDER BY avg(V0*I0) DESC)
SELECT * FROM ( SELECT DISTINCT ON (HOUSEID) * FROM hourlies ORDER BY HOUSEID ) q
ORDER BY q.houseid;
---
Q4. Top ten, 5 minute periods of consumption from house 'H1' (tumbling window).
WITH fives(houseid, avgpower) AS (
  SELECT HOUSEID, avg(V0*10) FROM POWER
  WHERE HOUSEID='H1'
  GROUP BY HOUSEID, (cast(EXTRACT(EPOCH FROM DATETIME) as int)/300)
  ORDER BY HOUSEID  ) 
SELECT fives.houseid, avgpower FROM fives ORDER BY fives.avgpower DESC LIMIT 10;
---
Q5. Number of samples between two datetimes
SELECT count(*) FROM POWER WHERE DATETIME >= '2012-07-30 09:35:00' 
AND DATETIME <= '2012-07-30 09:39:00';
---
Q6. Average maxium weekday consumption for each house.
WITH weekday_max(houseid, wdaymax, power) AS (
         SELECT HOUSEID, EXTRACT(dow FROM DATETIME), max(V0*I0)
         FROM POWER GROUP BY HOUSEID, EXTRACT(dow FROM DATETIME)
        )
SELECT houseid, avg(power) FROM weekday_max GROUP BY houseid ORDER BY houseid;
---
Q7. Number of samples taken between 5pm and 9pm on Wednesday.
SELECT count(*) FROM POWER
WHERE EXTRACT(dow FROM DATETIME) = 3 AND EXTRACT(hour FROM DATETIME) >= 17
AND EXTRACT(hour FROM DATETIME) <= 20;
\end{verbatim}
}

\section{LTSS Energy Data Benchmark Queries}
\label{sec:tssenergyqueries}
{\scriptsize 
\begin{verbatim}
Q1. Hourly average power consumption for house H1.
SELECT CTIME_hour, avg(V0*I0) FROM POWER WHERE HOUSEID = 'H1'
GROUP BY CTIME_hour ORDER BY CTIME_hour;
---
Q2. Maximum power sample for each house.
SELECT HOUSEID, max(V0*I0) FROM POWER
WHERE CTIME_hour > 8 AND CTIME_hour < 20 GROUP BY HOUSEID ORDER BY HOUSEID;
---
Q3. Highest hourly average power sample point for each house.
WITH hourlies (HOUSEID, HOUR, POWER) AS (SELECT HOUSEID, CTIME_hour, avg(V0*I0) 
     FROM POWER GROUP BY HOUSEID, CTime_hour ORDER BY avg(V0*I0) DESC)
SELECT HOUSEID, HOUR, max(POWER) FROM hourlies
WHERE HOUSEID IN (SELECT DISTINCT HOUSEID FROM POWER) GROUP BY HOUSEID;
---
Q4. Top ten, 5 minute periods of consumption from house 'H1' (tumbling window).
SELECT HOUSEID, avg(V0*I0), (TIMESTAMP / 300000000) FROM POWER 
WHERE HOUSEID='H1' GROUP BY HOUSEID, (TIMESTAMP / 300000000) 
ORDER BY (TIMESTAMP / 300000000) DESC LIMIT 10;
---
Q5. Number of samples between two datetimes
SELECT count(*) FROM POWER WHERE CTIME_year = 12 AND CTIME_month = 7 
AND CTIME_day = 30 AND CTIME_hour = 9 AND CTIME_min >= 35 AND CTIME_min < 39;
---
Q6. Average maxium weekday consumption for each house.
WITH weekday_max(houseid, wdaymax, power) AS (
         SELECT HOUSEID, CTIME_wday, max(V0*I0)
         FROM POWER GROUP BY HOUSEID, CTIME_wday)
SELECT houseid, avg(power) FROM weekday_max
GROUP BY houseid ORDER BY houseid;
---
Q7. Number of samples taken between 5pm and 9pm on Wednesday.
SELECT count(*) FROM POWER WHERE CTIME_wday = 3 AND CTIME_hour >= 17 
AND CTIME_hour <= 20;
\end{verbatim}
}





\end{appendix}

\end{document}